\documentclass[twocolumn,showpacs,preprintnumbers,amsmath,amssymb,aps]{revtex4}
\usepackage{graphicx}              
\usepackage{color}
\usepackage{amssymb}

\begin{document}

\title{Flows of R\'{e}nyi entropies}

\author{Yu. V. Nazarov}
\affiliation{Kavli Institute of NanoScience, Delft University of Technology, Lorentzweg 1, 2628 CJ, Delft, The Netherlands.}
\date{14-8-2011}


\begin{abstract}
We demonstrate that the condensed matter quantum systems encompassing two reservoirs connected by a junction permit a natural definition of flows of conserved measures, R\'{e}nyi entropies. Such flows are similar to the flows of physical conserved quantities such as charge and energy.
We develop a perturbation technique that permits efficient computation of R\'{e}nyi entropy flows and analyze second- and fourth order contributions.  Second-order approximation was shown to correspond directly to the transition events in the system and thereby to posess a set of "intuitive" features. The analysis of fourth-order corrections reveals a more complicated picture: the "intuitive" relations do not hold anymore, and
the corrections exhibit divergencies in low-temperature limit manifesting an intriguing non-analytical dependence of the flows on coupling strength in the limit of weak couplings and vanishing temperatures.
\end{abstract}

\maketitle
\section{Introduction}
Modern condensed matter theory borrows concepts from quantum information theory, n particular, entanglement. There is  an explosive growth of applications of  these conceps \cite{review}. Although the measures of entanglement are not linear in density matrix and as such cannot be directly related to quantum observables, these measures proved to provide insight into complex structure of many-body wave functions and in many cases can be efficiently computed numerically. \cite{Arovas}
Most applications are restricted to an equilibrium situation where the object of interest is the wave function of the ground state of a system $|\Psi>$.  The system is cut in to two subsystems $A,B$ (typically, in real space). Partial trace over a subsystem gives a reduced density matrix $\hat{R}^{(A)} ={\rm Tr}_B\lbrace|\Psi><\Psi|\rbrace$ in Hilbert space of another subsystem. Roughly, the von Neumann entropy of $\hat{R}$ can be ascribed to the boundary between the subsystems.
There is a number of illuminating theoretical results \cite{Calabrese,review,Fradkin} related to the von Neumann entropy of the boundary  and its scaling with the boundary area.
Not only entropy is under scrunity: the attention is paid to and the information is gathered from the whole spectum of the reduced density matrix $\hat{R}$.\cite{Arovas,Cramer,Song} The 
information about the spectrum can be compactified to 
a set of quantities that have been introduced in the context of information theory by Alfred R\'{e}nyi \cite{Renyi} and are called {\it R\'{e}nyi entropies}. We define them as follows: \cite{Definition}
$$
S_M = {\rm Tr} \left\lbrace \hat{R}^M\right\rbrace
$$
Logs of $S_M$ are obviously extensive quantities: if $\hat{R}$ can be presented as a direct product over a number of subsystems, their contrubutions to $\ln S_M$ add.  
The  von Neumann entropy is obtained by taking a formal limit
$$
S = -\lim_{M\to 1} \frac{\partial S_M}{\partial M} = -\lim_{M\to 1} \frac{\partial \ln S_M}{\partial M}. 
$$

Several works address time evolution of entropies in the course of transition processes like a quantum quench.\cite{quantumquench}.  There are interesting results concerning von Neumann entropy  production
in stationary non-equilibrium systems,  in particular, in quantum point contacts \cite{Beenakker}.
A simplest quantum point contact is a single-channel conductor with transmission coefficient $T_0$, that connects two electronic reservoirs. The entropy flow for zero temperature in the reservoirs  was shown to be \cite{Beenakker}
\begin{equation}
\frac{d S}{dt} = \frac{d N_{{\rm att}}}{d t} \left(T_0 \ln T_0 + (1-T_0) \ln (1-T_0)\right).
\label{eq:Beenakker}
\end{equation}
Here, $(d N_{{\rm att}}/dt) $ is the number of electrons per unit time that attempt to transfer the contact, $(d N_{{\rm att}}/dt) = eV/(2 \pi \hbar)$ The energy flow to either reservoir is given by $dE/dt = (dN_{{\rm att}}/dt) T_0 eV/2$. The approach of \cite{Beenakker} was rather heuristic relying on the representation of the electron many-body state  in terms of a sequence of individual single-electron scattering events \cite{LevitovFCS}, that is instrumental in the field of full counting statistics. Levitov and Klich have rederived these results with more microscopic approach, extented those to the case of a general time-dependent scatterer and discovered a remarkable correspondence between the entropy flow and full counting statistics of electron transfers. This has been further elaborated in \cite{KlichLeHur}.

There are two interesting peculiarities in the relation (\ref{eq:Beenakker}) that have not received a proper discussion so far. Firstly, non-analiticity of the expression (\ref{eq:Beenakker}) at small $T_0$ should indicate an intriguing divergence of perturbation series in $T_0$ that measures coupling strength between the reservoirs. Secondly, second law of thermodynamics relates the entropy and heat increments for a system in thermal equilibrium at temperature $T$. From this one would conjecture the relation between entropy and energy flows,
\begin{equation}
\label{eq:textbook}
\frac{d S}{dt} = \frac{1}{k_B T} \frac{d E}{dt},
\end{equation}
that would seem to have a textbook status. However, the result (\ref{eq:Beenakker}) is not compatible with this conjecture. The initial motivation for the research presented in this Article was to understand these peculiarities in a general calculable framework rather than for a restrictive case of quantum point contact.

The main result of the Article is that the flows of R\'{e}nyi entropies (Re-flows in short) are well-defined in standard condensed matter setups comprising two reservoirs connected by a junction. Importantly, they permit a detailed evaluation in the framework of a quantum perturbation theory developed here. This is in contrast to the entropy flow: albeit the latter can be readibly obtained by analytical continuation from integer to continous $M$.  We study second and fourth order of perturbation theory. The second-order results were shown to be very special: they satisfy the "textbook" relation (\ref{eq:textbook}) as well as some extra "intuitive" relations outlined in the text. This is because the second-order results can be put into direct correspondence with the transition events taking place in the system. Neither this correspondence nor "intuitive" relations hold for forth-order corrections. These corrections exhibit non-analytical low-temperature behavior manifesting non-analytical dependence of the flows on coupling strength in the limit of weak couplings and vanishing temperatures.

An obvious point of critisism is that the Re-flows are "unphysical" since they are non-linear in density matrix and  therefore cannot be readily asscociated with (the flows of) measurable quantities. The author shares the opinion and idea of Levitov and Klich \cite{KlichLevitov} that the flows of quantum infromation quantities may be related to the statistics of flows of quantum observables by universal relations whose general form is yet to discover. If this is true, a measurement of such statistics would in fact constitute a measurement of an "unphysical" flow. In any case,  the Re-flows characterize information exchange between the reservoirs in a rather detailed way. This makes their evaluation useful.

The structure of the paper is as follows. In Section \ref{sec:Measures}, we detail and illustrate the definition of R\'{e}nyi entropies. In Section \ref{sec:Flows}, we show that Re-flows can be associated with conserving currents, these currents being rather similar to electric or energy current through a junction in quantum transport setups. Next Section \ref{sec:perturbation} is devoted to the description of a generalized Keldysh technique that enables perturbative calculation of Re-flows in an arbitrary setup. We evaluate the flows in the secon-order approximation in Section \ref{sec:second}. The second-order approximation is characterized by a set of specific relations, those outlined
in Sections \ref{sec:fermigas}, \ref{sec:master}. In Section \ref{sec:fourth} we derive the fourth-order corrections. We demonstrate that  these corrections diverge in the limit of vanishing temperature (Section \ref{sec:divergence}) indicating non-analytical dependence of the flows in the limit of small couplings. 
We conclude in Section \ref{sec:conclusions}.

\section{Conserved measures}
\label{sec:Measures}
Let us discuss the definition and straightforward properties of R\'{e}nyi entropies. We start with an isolated finite quantum system 
that is characterized by density matrix $\hat{R}$.
We define the R\'{e}nyi entropies as traces of integer powers of $\hat{R}$,
\begin{equation}
S_M = {\rm Tr} \left\lbrace \hat{R}^M\right\rbrace
\label{eq:definition}
\end{equation} 
This definition can be easily extended  to non-integer $M$,
$$
S_M = {\rm Tr} \left\lbrace \exp\left( M \ln \hat R\right)\right\rbrace
$$
this identifies  $S_M$ as Laplace transform of the spectral function of
the operator $\ln \hat R$. This identification permits to relate the von Neumann entropy and R\'{e}nyi entropies,
\begin{equation}
S = - {\rm Tr} \lbrace \hat{R} \ln \hat R\rbrace = - \lim_{M\to 1} S_{M} = -\lim_{M\to 1} \ln S_M.
\end{equation}

If the system is in thermodynamic equilibrium at temperature $T$,
the R\'{e}nyi entropies are readily expressed in terms of the temperature-dependent free energy $F(T)$(e.g. \cite{Klimontovich}),
\begin{equation}
\ln S_M = \frac{M}{k_BT}\left(F(T/M) - F(T)\right).
\end{equation}
The R\'{e}nyi entropies can be easily computed for simple non-equilibrium quantum systems as well. The example we will use in this article concerns free fermions in a system of single-particle levels labelled by $k$. We ascribe the levels arbitraty filling factors $f_k$. The density matrix of the system reads
$$
\hat{R} = \prod_k \left( \tilde{f}_k(1-\hat{n}_k) +\hat{n}_k f_k\right),
$$
$\hat{n}_k \equiv \hat{a}^\dagger_k \hat{a}_k$ being number operator, $\hat{a}_k$ being annihilation operator in the level $k$, $\tilde{f}_k \equiv 1 -f_k$. From this, the R\'{e}nyi entropies read
\begin{equation}
\ln S_M =\sum_k \ln \left(\tilde{f}^M_k + f^M_k\right).
\label{eq:noneqFermi}
\end{equation}

For a degenerate Fermi gas in thermal equilibtrium with constant density of states $\delta_S^{-1}$ near the Fermi energy,  this reduces to
\begin{equation}
\ln S_M = \frac{\pi^2 k_BT}{6 \delta_S} \left(\frac{1}{M} -M\right).
\label{eq:ThermalFermi}
\end{equation}

The quantum evolution of the system is governed by a Hamiltonian $\hat{H}$. The same Hamitonian determines the evolution of the density matrix,
$$
-i \hbar \frac{d \hat{R}}{dt} = [\hat{H},\hat{R}].
$$ 
Since the density matrices in different momets of time are related by unitary transform, the trace of any power of $\hat{R}$ does not depend on time.

We just proved that R\'{e}nyi entropies provide a set of {\it conserved measures} for density matrix of an isolated quantum system,
$$
\frac{d}{dt} S_M =0,
$$
in similarity to the conserved physical quantities like energy or change. The difference is that $S_M$ are not linear in density matrix and therefore cannot be immediately associated with any quantum observables. 

While these definitions and properties may seem straightforward, there are some caveats to discuss. They are related to the traditional classical definition of entropy and its use. To illustrate, let us start with the non-equilibrium Fermi system in a pure state where $f_k$ are either $1$ or $0$. Obviously, $S_M=1$ for this pure state. Let us take $\hat{H}$ that involves weak interaction between the fermions. Common knowledge suggests that the interaction causes thermalization of the fermion distribution function. After some time, the filling factors $f_k$ will correspond to a Fermi distribution at some effective temperature $T^*$ that depends on the initial state. It may seem that the entropies are now given by Eq. \ref{eq:ThermalFermi} and thus differ from $1$. However, they must conserve! Indeed, after the Hamiltonian evolution the system is still in a pure state. 

The solution to this apparent paradox is that Eq. \ref{eq:noneqFermi} can not be used for an interacting Fermion system whatever small the interaction is. We could have used an alternative definition of the entopies that is close to the original classical definition of Boltzman. Namely, we could reduce the true density matrix $\hat{R}$ to $2\times2$ matrices $\rho_k$ by taking partial trace over all fermion Fock states not involving the level $k$, take the diagonal elements of these matrices and substitute them to \ref{eq:noneqFermi}. Such alternative definition might even seem more useful, for instance, for description of thermalization in an electron gas. However, this definition  would necessary involve entirely subjective choice of a basis (that of non-interacting fermion Fock states) and would not give rise to conveniently conserved measures. On these grounds, we stick to "quantum" definition (\ref{eq:definition}).

Let us now consider a bipartition of the Hilbert space $A\otimes B$ corresponding to two systems $A$ and $B$.
We can now define two sets of R\'{e}nyi entropies,
\begin{equation}
S^{(A)}_M = {\rm Tr}_A  \left\lbrace\left(\hat{R}^{(A)}\right)^M\right\rbrace; \;
S^{(B)}_M = {\rm Tr}_B  \left\lbrace\left(\hat{R}^{(B)}\right)^M\right\rbrace;
\end{equation}
where the reduced density matrices in two subspaces are defined 
with the aid of the partial traces in these subspaces,
\begin{equation}
\hat{R}^{(A)} = {\rm Tr}_B \lbrace \hat{R}\rbrace; \;
\hat{R}^{(B)} = {\rm Tr}_A \lbrace \hat{R}\rbrace.
\end{equation}
If the quantum evolutions of the systems are completely independent, 
$$
\hat{H} = \hat{H}_A + \hat{H}_B,
$$
$H_{A,B}$ being operators involving the corresponding subspaces only, both sets provide the conserved measures,
$$
\frac{d}{dt} S^{(A)}_M = \frac{d}{dt} S^{(B)}_M=0.
$$

It is interesting to note that the sets of R\'{e}nyi entropies are not the only conserved measures characteristic for a bipartition. Any polynomial in density matrix that is invariant with respect to the group $U_A \otimes U_B$ of unitary transforms in two subspaces, would provide such a measure. To give a minimal example, let us label the states in $A$($B$) with Latin (Greek) indeces. The quantity 
\begin{equation}
K \equiv \sum_{a,b,c;\alpha,\beta,\gamma} R_{a \alpha,b \gamma} R_{b \beta, c\alpha} R_{c \gamma, a \beta}  
\label{eq:other-measure}
\end{equation}
is a conserved measure that can be reduced neither to the R\'{e}nyi entropies of the systems nor to the R\'{e}nyi entropy of the whole system. The characterization of all such measures forms an interesting research task beyond the scope of this Article.

\section{Flows}
\label{sec:Flows}
Let us complicate the bipartition situation by including a Hamiltonian coupling between the systems $A$ and $B$,
\begin{equation}
\hat{H} = \hat{H}_A +\hat{H}_B +\hat{H}_{AB},
\end{equation}
$\hat{H}_{AB}$ being an operator that involves degrees of freedom in both subspaces. 

We will assume that the systems $A$ and $B$ are infitely large and are characterized by continuous excitation spectrum given by  $\hat{H}_{A,B}$ respectively. However, $H_{AB}$  couples a relatively small number of degrees of freedom in both systems. It is convenient to assume that the systems $A$, $B$are in thermal equilibrium at different temperatures $T_A,T_B$ and/or chemical potentials $\mu_A,\mu_B$ and thus play a role of reservoirs. The coupling is such that it does not lead to equilibration of $T_{A}, T_{B}$ ($\mu_A,\mu_B)$. Rather, there is a constant energy/particle flow between the systems that does not depend on actual volume of $A$ and $B$, nor on their detailed properties. 

\begin{figure}
\includegraphics[width=0.35\textwidth]{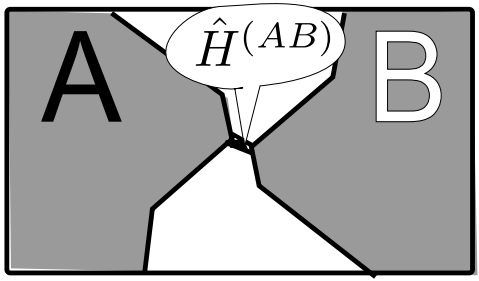}
\caption{ Two leads (reservoirs) $A$ and $B$ in a typical quantum transport setup correspond to the bipartition $A\otimes B$. The junction connecting the two is associated with coupling Hamiltonian $H^{(AB)}$. In similarity with the flows of charge and energy, one can define the flows of R\'{e}nyi entropies (Re-flows) in such setups.}
\label{fig:qtsetup}
\end{figure}

Examples of such arrangement are common quantum transport setups. \cite{QuantumTransport}
An exemplary setup (Fig. \ref{fig:qtsetup}) consists of two (infitite) current-carrying leads $A$ and $B$ kept at different chemical potentials. These leads are connected by a junction. Simplest coupling that we will use in our examples is the tunneling Hamiltonian
\begin{equation}
\hat{H}_{AB} = \sum_{k_A,k_B} t_{k_A,k_B} \left(\hat{a}^\dagger_{k_A} \hat{a}_{k_B} + {\rm h.c} \right)
\label{eq:tunnelHam}
\end{equation}
that describes particle transfer between the leads.

It is important in quantum transport setups that the value to measure experimentally and compute theoretically is a flow of a conserved quantity, electric charge being a simplest and most usefull example. In principle, one defines a flow of charge through a cross-section
of the setup, this cross-section may also define the bipartition. Owing to conservation, this flow does not depend on the cross-section. Therefore, the exact bipartition is also not important. Owing to this, the current is said to depend on the properties of the junction rather than on details of the reservoirs. This enables the experimental investigation of nanostructures: a signal in a measurement of a non-conserving quantity would be most probably dominated by massive leads rather than by a small nanostructure, while the flow of conserved charge is determined by the nanostructure forming the junction between the leads. The same pertains energy flow. Energy and charge are not the only conserved quatities: approximately conserved spin currents \cite{Spintronics} provide another practical example. Artificially constructed conserved quantities appear useful in description of quantum coherence in nanostructures, being mathematical basis of so-called circuit theory of quantum transport. \cite{QuantumTransport}

From analogy with the flows of conserved quantities in quantum transport setups, we conjecture that there are finite flows of conserved measures, R\'{e}nyi entropies (Re-flows), to each subsystem $A$ and $B$,
$$
\frac{d}{dt} \ln S^{(A),(B)}_M \equiv {\cal F}^{(A)}_M ;
$$
Owing to conservation of R\'{e}neyi entropy in each system, the Re-flows would not depend on exact bipartition of the system and are determined by properties of the junction that is in principle described by $\hat{H}_{AB}$.

There is an important difference. For physical quantities the conservation holds in the whole system as well as in each subsystems. For instance, elecrtical currents to each lead must satisfy  $I_A+I_B =0$.\cite{spin-currents}
 As far as R\'{e}nyi entropies are concerned, there is no exact conservation law for a sum
 $\ln S^{(A)}_M + \ln S^{(B)}_M $ at finite $\hat{H}_{AB}$, although these quatities are extensive. There is a conservation law for the total R\'{e}nyi entropy $\ln S^{(A+B)}$. However, the latter at finite $\hat{H}_{AB}$ is the sum  $\ln S^{(A)}_M + \ln S^{(B)}_M $ only approximately, up to the terms proportional to the volume of the system.
 
 Therefore, in general 
$$ 
{\cal F}^{(A)}_M + {\cal F}^{(B)}_M \ne 0.
$$ 

Let us compute these Re-flows. We will always restict ourselves to the Re-flow to the system $A$:
the Re-flow to the system $B$ is obtained by permutation of $A$ and $B$. For brevity, we  will skip the 
index $A$ in $S^{(A)}_M$ and ${\cal F}^{(A)}_M$ where this does not lead to confusion.
  
\section{Perturbation technique}
\label{sec:perturbation}
\begin{figure}
\includegraphics[width=0.35\textwidth]{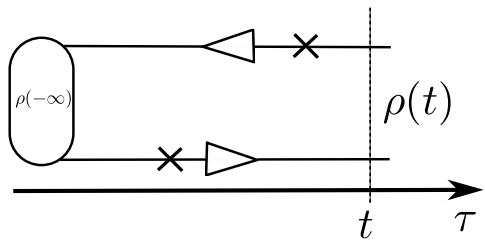}
\caption{ Perturbation theory for a single density matrix on the Keldysh contour.}
\label{fig:commonkeldysh}
\end{figure}

We will use a perturbation technique in $\hat{H}_{AB}$ and keep the calculation as general as possible. We assume "adiabatic switching" of the perturbation\cite{Landafshiz}:  far in the past the coupling is absent, and the density matrix is a direct product over subspaces,
\begin{eqnarray*}
\hat{R}(-\infty) = \hat{R}_A(-\infty) \otimes \hat{R}_B(-\infty) \\
\hat{R}_A(-\infty) = \sum_a p_a |a><a| ;\; \hat{R}_B(-\infty) = \sum_\alpha p_\alpha |\alpha><\alpha|.
\end{eqnarray*}
(As above, we label the states in subspaces $A$ ($B$) with Latin (Greek) indexes. )
The coupling slowly grows achieving actual values at time $t$. The time evolution is given by (from now on, $\hbar =1$)
\begin{eqnarray*}
\hat{R}(t)  = {\rm Texp}\left(i\int^t_{-\infty} d\tau \hat{H}_{AB}(\tau)\right) \hat{R}(-\infty) \times\\
\times {\rm \tilde{T}exp}\left(-i\int_{-\infty}^t d\tau \hat{H}_{AB}(\tau)\right)
\end{eqnarray*}
$\hat{H}_{AB}(\tau)$ is taken here in interaction representation, ${\rm Texp}({\rm \tilde{T}exp})$ denote time(anti)ordering in the evolution exponents. Expanding this in  $H_{AB}(\tau)$ gives perturbation series most conveniently presented as diagramms 
involving the Keldysh contour (Fig.\ref{fig:commonkeldysh}). The operators in perturbation series are ordered along the contour. Two parts of the contour correspond to time evolution of bra's and ket's in the density matrix. The crosses represent the (time-dependent) perturbation $H_{AB}(t)$ at a certain time moment. The intergatation over time moments of all perturbations is implied. There is a state index associated with each piece of the contour. Since $\hat{R}(-\infty)$ is diagonal, this index does not change when passing this element.
The index changes if a non-diagonal matrix element of the perturbation is considered. Summation over indices is implied.

\begin{figure}
\includegraphics[width=0.35\textwidth]{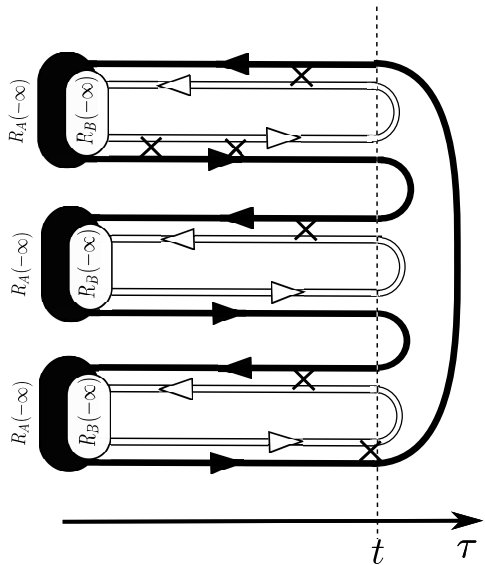}
\caption{ A diagramm of perturbation theory for $S^{(A)}_M$ for $M=3$. It involves
three parallel worlds. Reconnection of Keldysh contours for subspaces $A$(black) and $B$(white) accounts for partial trace over $B$ and matrix multiplication in $A$.}
\label{fig:3M}
\end{figure}

The crucial observation is that this scheme can be straightforwardly generalized to any integer number $M$ of density matrices. These matrices undergo independent unitary evolution in time interval $(-\infty,t)$.
It is constructive to think of a set of $M$ "parrallel worlds" and draw the diagramms for perturbation series  using $M$ parallel bra- and ket-contours.  To compute $S_M(t)$ with this set, we first need to 
'split' the contours to account for possibly different ordering of operators in subspaces $A$and $B$ (black and white curves in Fig. \ref{fig:3M}). Then we need to reconnect the countours at  $\tau = t$. All white contours are closed within each world, this corresponds to the partial trace over $B$ for each density matrix involved. In contrast to this, the black contours are connected to form a single loop going through all the worlds, this corresponds to the matrix multiplication in the definition (\ref{eq:definition}) of R\'{e}nyi entropy.  This conveniently represents  the rules of operator ordering for any diagram of particular order in $H_{AB}$.  

It is interesting to note that the reconnecting the contours in a different fashion gives rise to perturbation theory for other conserved measures, for instance, for $K$ given by Eq.{eq:other-measure}.

It is natural to require that the matrix elements  of $H_{AB}$ are only non-diagonal, that is, $H^{(AB)}_{a \alpha, b \beta} = 0$ if either $a=b$ or $\alpha=\beta$. 
In this case, the first non-vanishing contibutions to Re-flows will be of the second order in $H_{AB}$. 

\section{Second order}
\label{sec:second}
\begin{figure}
\label{fig:second-order}
\includegraphics[width=0.35\textwidth]{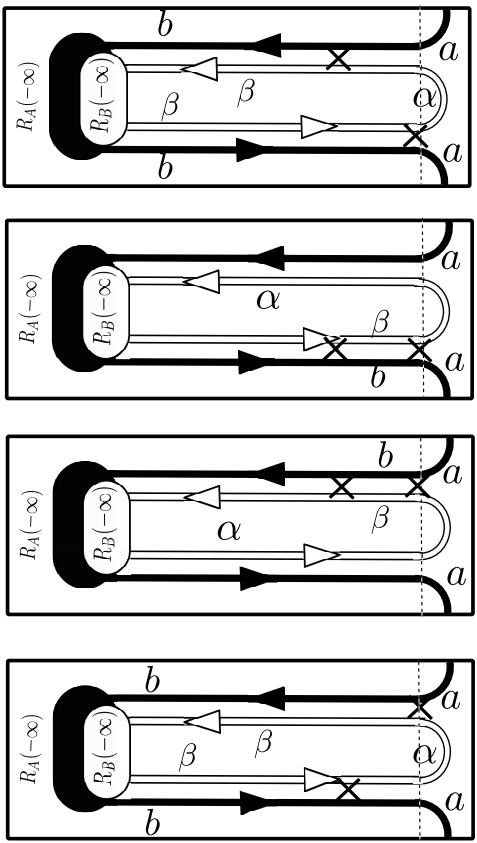}
\caption{ Second order diagramms for time derivative of a R\'{e}nyi entropy.
The contibutions come only from perturbations $\hat{H}^{(AB)}$ in
the same world, only this world is shown in each diagramm. For all diagramms, the perturbations are taken at time moments $t$ and $t' <t$.
The letters at the contours label the states involved.}
\end{figure}

Let us compute the Re-flows in the second order in $H_{AB}$.
It is proficient to directy compute the time-derivative of $S_M$. For diagramms, this corresponds to placing one of the perturbations at $\tau=t$. The only way to satistfy the continuity of state index along the white contours is to place the second perturbation in the same world. Four contibuting diagramms are given in Fig. \ref{fig:second-order}. In fact, the same four diagramms arise in the derivation of Golden Rule transition rate and are familiar to anyone who made use of Keldysh perturbation theory for density matrix. The specifics of R\'{e}nyi entropies is reflected in extra factors $p^{M-1}_a$ the diagramms acquire in comparison with the case of a single density matrix. Summing up the four diagramms yields 
\begin{eqnarray}
\frac{\partial}{\partial t} S_M= 
\left(-M \sum_{a,\alpha;b,\beta} |H^{(AB)}_{a\alpha,b\beta}|^2 p^M_ap_\alpha \right.\\ \nonumber
\left.+M \sum_{a,\alpha;b,\beta}  |H^{(AB)}_{a\alpha,b\beta}|^2 p_bp_\beta p^{M-1}_a \right) 
\\ \nonumber
\int_{-\infty}^{t} dt' 2 {\rm Re} \left( e^{i(t-t')(E_i+E_\alpha -E_j-E_\beta)}\right)   
\end{eqnarray}
The integral over time $t'$ reduces to 
$$
2\pi \delta(E_a+E_\alpha-E_b-E_\beta),
$$
manifesting energy conservation between the initial state $|a\alpha>$ and final state$|b\beta>$.

This suggests that we can rewrite the whole expression in terms of Golden Rule 
rates $\Gamma_{a\alpha,b\beta}$ of the transitions between the states $|a\alpha>$
and $|b\beta>$,
\begin{equation}
\Gamma_{a\alpha,b\beta} = 2\pi |H^{(AB)}_{a\alpha,b\beta}|^2\delta(E_a+E_\alpha-E_b-E_\beta).
\end{equation}

With this, the flow reads
\begin{equation}
\left(S_M\right){\cal F}_M = 
M \sum_{a,\alpha;b,\beta} \Gamma_{a\alpha;b\beta}(p_b p_\beta-p_a p_\alpha)p^{M-1}_a 
\label{eq:2nd-general-form1}
\end{equation}
We see that the flow {\it vanishes} if the systems are in thermodynamic equilibrium at the same temperature. Indeed, in this case $p_b p_\beta / p_a p_\alpha = \exp((E_b+E_\beta-E_a-E_\alpha)/k_BT)=1$. 

Since the transition rates $\Gamma_{a\alpha,b\beta}$ in Golden rule approxiamtion are symmetric with respect to a permutation $a\alpha \leftrightarrow b\beta$, we can regroup the terms  to arrive at
\begin{equation}
\left(S_M\right){\cal F}_M =  
M \sum_{a,b} \Gamma_{a\to b}p_a(p^{M-1}_b-p^{M-1}_a). 
\label{eq:2nd-general-form2}
\end{equation}
where 
$$
\Gamma_{a \to b} = \sum_{\alpha,\beta}  \Gamma_{a\alpha;b\beta}p_\alpha 
$$ 
gives the total transition rate from the state $|a>$ to the state $|b>$ averaged over all possible configurations of system $B$. Let us use Eq. \ref{eq:2nd-general-form2} to derive a simplified expression valid in zero-temperature limit. In this limit, the system $A$ is initially in the ground state $|0>$, so that $p_0=1$ and $p_a=0$ for $a \ne 0$, $S_M =1$.
We obtain 
\begin{equation}
{\cal F}_M = - M \Gamma_0;
\label{eq:2ndorder-zeroT}
\end{equation}
$\Gamma_0$ being the total transition rate from the ground state to any other state.
Remarkably, this involves no assumption concerning the system $B$: it can be very far from equilibrium.

Eq.  \ref{eq:2nd-general-form2} is also a convenient starting point to derive the expression for the von Neumann entropy flow. Taking the limit $M \to 1$, we obtain 
\begin{equation}
-\frac{\partial S}{\partial t} =  \sum_{a,b} \ln\left(p_b/p_a\right)\Gamma_{a \to b} p_a. 
\label{eq:2nd-entropy}
\end{equation}
Let us assume thermal equilibrium of $A$. In this case, $\ln\left(p_b/p_a\right) = (E_a -E_b)/k_BT)$. Summing up the energy changes $E_b -E_a$ in the course of individual transitions from $a$  to $b$, we prove that the energy flow to the system $A$ equals
$$
\frac{d E}{d t} = \sum_{a,b} \Gamma_{a \to b} (E_b-E_a) p_a
$$  
Comparing this with Eq. \ref{eq:2nd-entropy}, we recover the relation (\ref{eq:textbook}) that appears to be universaly valid within the second-order perturbation theory. Remarkably, this invloves no assumption about the system $B$.

One can easily imagine the situation where $dE/dt$ remains constant at $T\to 0$ (see Eq. \ref{eq:Beenakker}). The entropy flow in this case diverges as $1/T$. One can wonder how it is compatible with the fact that Re-flows approach finite limit at vanishing temperature.
The point is a hidden non-analyticity of (\ref{eq:2ndorder-zeroT}): the flow does not vanish in the limit $M\to 1$ as it is implied by more general Eq. \ref{eq:2nd-general-form2}. Therefore, Eq. \ref{eq:2ndorder-zeroT} cannot be used to derive the flow of von Neumann entropy.

\section{Example: Fermi gas}
\label{sec:fermigas}
Let us explicitly compute the Re-flows for tunneling between 
two non-equibrium Fermi gases. 
In this case, $a$ (or $b$) is a number state defined by a set of occupation numbers in all levels $k$ $\{ n_k\}$,
$p_a = \prod_k (\tilde{f}_k (1-n_k) + f_k(n_k))$. The perturbation $\hat{H}_{AB}$ is given by Eq. \ref{eq:tunnelHam}.
Let us concentrate on the transitions involving a level $q$.  Since the probabilities factorize, we can forget about the Fock states involving all other levels since their contributions to right hand side of Eq. \ref{eq:2nd-general-form2} cancels with their contribution to $S_M$. We can therefore consider only two states spanned by level $q$, subsituting into Eq. \ref{eq:2nd-general-form2}
their contribution to $S_M = f^M_q +\tilde{f}^M_q$. 
The probablilities of these states, $|0>$ ($n_q  = 0$) and $|1>$ ($n_q=1$) are $f_q$ and $\tilde{f}_q$ respectively. Substituting this into Eq. \ref{eq:2nd-general-form2} and summing up over $q$ yields
 \begin{equation}
\label{eq:2nd_neFermi_q}
 {\cal F}_M =
M \sum_{q} \left(\Gamma^+_{q} \tilde{f}_q -\Gamma^{-}_q f_q\right) \frac{f^{M-1}_q -\tilde{f}^{M-1}_q}{f^{M}_q +\tilde{f}^{M}_q}, 
\end{equation}
$\Gamma^{+}_q$ ($\Gamma^{-}_q$) being the rates of addition (extraction) of a fermion to (from) the level $q$.

Let us outline a heuristic way to obtiain the above relation. This way does not involve any quantum mechanics: it is something L. Boltzmann would do if given the problem. Let us treat $\ln S_M$ and $f_q$ as continuous quantities related by Eq. \ref{eq:noneqFermi}. By virtue of the relation, the flux of $\ln S_M$ would be expressed in terms of fluxes of $f_q$ as
\begin{equation}
\frac{d}{dt} \ln S_M = \sum_q \frac{\partial \ln S_M}{\partial f_k} \left(\frac{d f_k}{d t}\right).
\label{eq:boltzmann}
\end{equation}
Since $f_q$ is the average number of particles in the level $q$,
$$
\frac{d f_q}{d t} = \Gamma^+_{q} \tilde{f}_q -\Gamma^{-}_q f_q
$$
Combining these two relations reproduces Eq. \ref{eq:2nd_neFermi_q}! We stress that we do not see any reason for this relation to hold in general, since both $S_M$ and $n_q$ change in discrete rather than continuous fashion in the course of a tunneling transition. Indeed, the inspection of forth-order terms in perturbation theory shows that the heuristic way does not generally work.

Let us assume that the filling factors in the system $A$ depend on single-particle energy $E$ only. Eq. \label{eq:2nd-neFermi-q}
reduces to \begin{equation}
{\cal F}_M =
M \int dE \left(\Gamma^+\tilde{f}-\Gamma^{-}f\right) \frac{f^{M-1} -\tilde{f}^{M-1}}{f^{M} +\tilde{f}^{M}},
\label{eq:2nd-neFermi-E}
\end{equation}
$\Gamma^{+,-}(E)$ being addition/extraction rates per energy interval.
In case of vanishing temperature (in the system $A$), $f_q = \Theta(-E+\mu)$. The R\'{e}nyi entropy flow reduces to 
 \begin{eqnarray*}
{\cal F}_M = M \int dE \left( \Gamma^+(E) \Theta(E-\mu) \right.\\
\left. -\Gamma^-(E) \Theta(-E+\mu) \right) 
{\rm sgn}(E-\mu) =\\
= - M \frac{d}{dt} (N^+ +N^-)
 \end{eqnarray*}
 $N^{\pm}$ being numbers of added/extracted particles.  This is in agreement with the general relation (\ref{eq:2ndorder-zeroT})
 since the  only events that happen in our setup are indeed additions or extractions of single particles.
 As in general case, analytical continuation to $M\to1$ at vanishing temperature does not work since the expression does not vanish in the limit  $M\to 1$.  One can work at non-zero temperature, where it does vanish. Taking derivative of Eq. \ref{eq:2nd-neFermi-E}
at $M\to 1$ yields the flow of von Neumann entropy
$$
\frac{\partial S}{\partial t} = \int dE \left( \Gamma^+(E) \tilde{f}(E)  -\Gamma^-(E) f(E) \right) \ln\left(\frac{f}{\tilde{f}}\right) 
$$
Substituting Fermi distribution, we recover once again the relation (\ref{eq:textbook}) between the entropy and energy flows.

\section{Master equation approach}
\label{sec:master}
\begin{figure}
\includegraphics[width=0.35\textwidth]{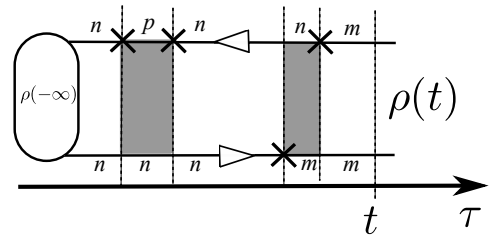}
\caption{ Master equation in a single world is obtained from the resummation of perturbation series whereby the time-line is separated into diagonal and non-diagonal (grey-shaded) blocks. One can do this resummation in multiple worlds to compute the time evolution of R\'{e}nyi entropies.}
\label{fig:masterequation}
\end{figure}

In a single world, there is a simple way to resum the perturbation series and arrive to a master equation that contains only diagonal elements of density matrix (Fig. \ref{fig:masterequation}). For a diagramm, we split the time-line by perturbations into the blocks
as shown in the Figure. The blocks come in two sorts: diagonal with the same state index in both contours  and non-diagonal
ones. Intergation over time duration of non-diagonal blocks gives Golden Rule transition rates. The integration over time duration of diagonal blocks is divergent, this indicates that the diagramms need to be resummed. The resummation gives the master equation invoving the diagonal elements of the density matrix only. In case of bipartition, this equation reads

\begin{equation}
\frac{d}{dt} p_{a \alpha} = - \Gamma_{a\alpha} p_{a \alpha} + \Gamma_{a\alpha,b\beta} p_{b\beta}.
\end{equation}
Here $\Gamma_{a\alpha}$ stands for the total transition rate from the state
$|a\alpha>$, that is the sum of the partial transition rates, $\Gamma_{a\alpha}=\sum_{b\beta} \Gamma_{a\alpha,b\beta}$.  

Inspired by the above exercises with second-order diagrams, one can attempt the resummation in parallel worls in such a way that the perturbations in a non-diagonal block are situated in the same world. In this case, we have independent master equations governing the dynamics of density matrix in each world. 

Suppose we know the propagator of the master equation that is, a linear kernel  ${\cal P}$ expressing the evolution of the probabilities, 
$$
p_{a\alpha}(t) = {\cal P}_{a\alpha,b\beta}(t,t') p_{b\beta}(t').
$$

With this, we can easily express time evolution of $S_M$ in time interval $(0,t)$
\begin{equation}
S_M(t) = \sum_a \left(\sum_\alpha {\cal P}_{a\alpha,b\beta}(t,0) p_b(0)p_\beta(0)\right)^M
\end{equation}
and compute the R\'{e}neyi entropy flows. We will not analyze the general situation here since as all systems described by a master equation  it belongs to the realm of classical physics anyway. Rather, we concentrate on a simple but constructive example. We will assume that initially the systems were in a pure product state. Let us denote this state $|00>$. Let us further assume that the transitions from this state as well from all other state can happen to a great number of states. This number is so big that once the  first transition has taken place the system will never get back to the original state.
With this, the propability to remain in the state $|00>$ falls off exponentially,
$$
p_{00} = \exp\left(-\Gamma_{00} t\right),
$$
$\Gamma_{00}$ being total rate of all transitions from $|00>$. Under same conditions, the sum in $S_M$ is dominated by the probability to remain in the state $|00>$. The $S_M(t)$ just equals the probability that during this time interval no transition event occurs in any of $M$ parallel worlds, 
\begin{equation}
S_M(t)  = p^{M}_{00}=\exp \left(-M\Gamma_{00}t\right).
\label{eq:primitive}
\end{equation}
The flow is therefore ${\cal F}_M = - M \Gamma_{00}$, in full correspondence with Eq. \ref{eq:2ndorder-zeroT}.

We note that the resummation of the diagramms we have done is not complete since we have assumed that each non-diagonal block comprises only two perturbations. The transition rates are thus approximated by their Golden Rule expressions. In most cases involving master equation in a single world, this is not important: the further orders of perturbation theory provide small  corrections to the rates not changing the structure of the equation. It looks like this is not so in the case of parallel worlds.
The analysis of fourth-order perturbation corrections  presented below indicates the problems invloved. 
 
\section{Fourth order}
\label{sec:fourth}

Let us analyze the fourth-order diagramms for time derivative of $S_M$. As above, we assume that $H_{AB}$ does not contain diagonal elements. Since white contours are closed within each world, the four perturbations can either all come in the same world or in two pairs in two different worlds. If all four come in the same world, they describe a correction to one of the Golden Rule transition rates. This correction does not bring anything new and we disregard these diagramms in further consideration. 

\begin{figure}
\label{fig:fourth-order}
\includegraphics[width=0.35\textwidth]{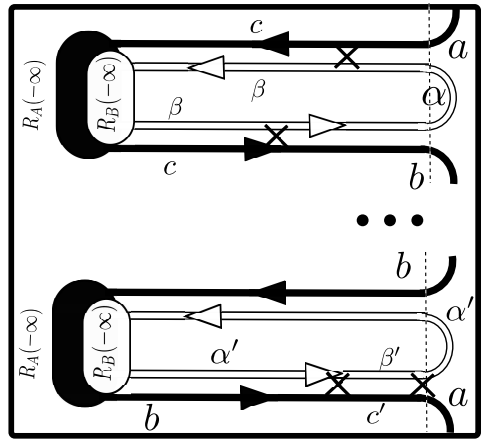}
\caption{ A fourth-order "quantum" diagramm for R\'{e}nyi entropy flows.
The contibutions come from perturbations $\hat{H}^{(AB)}$ in
two different worlds, only these two worlds are shown. The letters on the contours label the states involved.}
\end{figure}

A diagram involving two different worlds is given in Fig. \ref{fig:fourth-order}. We see that in general the black contour entering a world with perturbations exits it with a different state index. For a particular case when these indices are the same,   
$a =b$, the diagramm diverges upon integration over time. This is not surprising since  we expand $S_M(t) \propto \exp ( {\cal F}_M t)$. The fourth-order expansion thus contains terms $\propto ({\cal F}^{(2)}_M)^{2} t/2$, ${\cal F}^{(2)}$ being the second-order contibution to the rate that we have already calculated. Indeed, the diagram with $a=b$ is proportional to $({\cal F}^{(2)})^2$ and therefore does not contibute to fourth-order correction to the flow. 
We thus concentrate on the case $a \ne b$. We call this diagram "quantum" since we will see that it does not permit an interpretation in terms of "classical" transition events.
All expressions for ${\cal F},dS/dt$ in this Section give fourth-order corrections to these quantities. 

There are 16 diagramms of this sort corresponding to the number of ways the pairs of $\hat{H}^{(AB)}$ in each world can be placed on bra and ket contours.
Summing up all of them, we can present the fourth-order correction in the following form:
\begin{eqnarray}
\label{eq:4th-general}
\frac{d}{dt} S^{(A)}_M = \pi \sum_{a,b} |A_{ab}|^2 \delta(E_a -E_b)\frac{p^{M-1}_a-p^{M-1}_b}{p_a-p_b};\\
A_{ab} = \sum_{c,\alpha,\beta} H^{(AB)}_{a\alpha,c\beta} H^{(AB)}_{c\beta,b\alpha} \nonumber\\
\left(\pi \left((p_a+p_b)p_\alpha - 2 p_cp_\beta\right) \delta\left(E_a+E_\alpha -E_c-E_\beta\right) \right. \nonumber \\
\left.- i\frac{p_a -p_b}{E_a+E_\alpha -E_c-E_\beta}\right).
\nonumber
\end{eqnarray}
The structure of the matrix elements in the "amplitude" $A_{ab}$ is the same as for an amplitude of the transition from the state $|a\alpha>$ to the state $|b\alpha>$, that is, without the change of the state of the subsystem $B$. Such transition would seem to involve a virtual state $|c,\beta>$. However, the rest of the expression for $A_{ab}$ does not support this interpretation: rather, probabilities enter in a form suggesting that the transition takes place between one of the states $|a\alpha>,|b\alpha>$ and the state $|c\beta>$. Therefore, the expression can be associated with no "classical" transition and corresponds to no actual transition rate.

Let us assume that the probabilities in the system $A$ depend only on energies of the corresponding states. Then it follows from $E_a=E_b$ that $p_a=p_b$. The term in $A_{ab}$ with the energy difference in the denominator vanishes and the flow reduces to 
\begin{eqnarray}
\label{eq:4th-f-of-E}
 S_M  {\cal F}_M= (M-1)\pi \sum_{a,b} |A_{ab}|^2 \delta(E_a -E_b) p^{M-2}_a;\\
A_{ab} = 2\pi \sum_{c,\alpha,\beta} H^{(AB)}_{a\alpha,c\beta} H^{(AB)}_{c\beta,b\alpha} \nonumber\\
\left(p_a p_\alpha - p_c p_\beta\right) \delta\left(E_a+E_\alpha -E_c-E_\beta\right). \nonumber 
\end{eqnarray}
We notice that if both systems are in thermal equilibrium, it follows from $E_a+E_\alpha =E_c+E_\beta$ that 
$p_a p_\alpha - p_c p_\beta$ and the "amplitudes" $A_{ab}$ vanish.

Let us compute the fourth-order contribution for non-equilibrium Fermi gas. We pick up  two levels in system $A$, $k_{1,2}$, with $E_{k_1} =E_{k_2} = E$. Their contibution to $S_M$ reads $S_M = (f_A(E) + \tilde{f}_A(E))^2$. Let the state $a$ correspond
to $|1>_{k_1}|0>_{k_2}$ and state $b$ correspond to $|0>_{k_1}|1>_{k_2}$:  The probablilities of these states equal $f_A(E)\tilde{f}_A(E)$.  The "amplitude" $A_{ab}$ is contributed by all levels $q$ in the system $B$ with the energy $E_q=E$. There are two possible "virtual states"
$|c\beta>$. 
First possibility is an empty level $q$ and both levels $k_{1,2}$ occupied ($p\alpha = f_B(E)$, $p_\beta =\tilde{f}_B(E)$, $p_c =f^2_A(E)$). 
Second possibility is an occupied level $q$ and empty levels $k_{1,2}$($p\alpha = \tilde{f}_B(E)$, $p_\beta =f_B(E)$, $p_c =\tilde{f}^2_A(E)$). 
Owing to anticommutativity of fermionic creation/annihilation operators, these two contibutions to the "amplitude" come with opposite sign yielding 
\begin{equation}
A_{ab} = 2\pi (f_B(E)-f_A(E)) \sum_q t_{k_1,q} t_{k_2,q} \delta(E_q -E). 
\label{eq:4th-Fermi-amplitude}
\end{equation}
To get the contibution of $k_{1,2}$ to the Re-flow, we square $A_{ab}$, multiply it with $p^{M-2}_\alpha =(f_A\tilde{f}_A)^{M_2}$, and divide with the contribution $S_M = (f^M_A+\tilde{f}^M_A)^2$ of these levels to R\'{e}nyi entropy. Next to it, we sum over all states in the energy interval $(E,E+dE)$ and integrate over energy to obtain the total Re-flow. We obtain
\begin{equation}
\label{eq:4th-Fermi-answer}
{\cal F}_M = (1-M) \int dE (f_B-f_A)^2 \frac{(f_A\tilde{f}_A)^{M-2}}{(f_A^M+\tilde{f}_A^M)^2} \Xi(E);
\end{equation}
where $\Xi(E)$ is defined as
\begin{eqnarray*}
\Xi(E) = 8\pi^2 \sum_{k,k',q,q'} t_{k,q}t_{k',q'}t_{k',q}t_{k,q'} \\
\times  \delta(E-E_k) \delta(E-E_k') \delta(E-E_q) \delta(E-E_q').
\end{eqnarray*}

\section{Low-temperature divergencies}
\label{sec:divergence}
We will now demonstrate that the "quantum" contibution derived manifests serious problems with term-by-term perturbation theory in the limit of vanishing temperature, indicating non-analytical dependence of the flows on coupling strength in the limit of weak couplings and vanishing temperatures.

First let us note that the contibution seems to have an evident zero-temperature limit, namely zero, at least if the ground state of the system $A$ is not degenerate. Indeed, delta-funcion in Eq. \ref{eq:4th-general} cannot be satisfied for any state $b\ne a$. However, analythical continuation to non-integer $M$ gives rise to problems. Let is illustrate this computing (\ref{eq:4th-Fermi-answer}) assuming thermal distribution and the difference between chemical potentials in $A$ and $B$, $eV$, is much bigger than $k_BT$. To this end, we can set $f_B=0$ in Eq. Disgerarging the energy dependence of $\Xi$ near Fermi surface, we obtain
\begin{equation}
{\cal F}_M = k_BT\Xi \frac{1-M}{M} \left(1+\frac{2\pi}{M}
\left(
\frac{1}{\sin\frac{\pi}{M}}
-\frac{1}{\sin\frac{2\pi}{M}}
\right)\right)
\end{equation}
While the flow vanishes at vanishing temperature, it has a finite limit ${\cal F}_1 = 3 k_B T \Xi$ at $M\to 1$ where it should vanish. 

This signals the divergence of fourth-order contribution to the flow of von Neumann entropy. Indeed, from the general expression (\ref{eq:4th-f-of-E}) we deduce that
\begin{equation}
\frac{d S}{dt} = \sum_{a,b} |A_{ab}|^2 \delta(E_a -E_b)\frac{1}{p_a};
\end{equation}
that is, the states with lesser probabilities $p_a$ contibute most to the entropy flow! For Fermi gas example,
the entropy flow reduces to
\begin{equation}
\frac{d S}{dt} =\Xi \ \int dE \frac{(f_A-f_B)^2}{f_A \tilde{f}_A}
\end{equation}
For equal temperatures in the systems $A$ and $B$, this gives
\begin{equation}
\frac{d S}{dt} = 4 k_B T\  \Xi \ \sinh^2 \frac{eV}{2 k_BT}.
\end{equation}
The fourth-order contibution to the entropy flow thus diverges exponentally upon $T\to 0$.
\section{Conclusions}
\label{sec:conclusions}
In this paper, we have shown that the flows of R\'{e}nyi entropies are well-defined for common setups encompassing a junction between two reservoirs $A$ and $B$. They are similar to flows of physical conserved quanities like energy and charge. 
They vanish in thermal equilibrium and can be induced by the difference of chemical potentials/temperatues in the reservoirs.
The flows characterize (quantum) information transfer between the reservoirs and generalize the flow of von Neumann entropy.

We develop a general quantum perturbation technique to compute these flows, this involves a set of parallel worlds.  We consider second and forth order terms. In second-order approximation, the flows can be related to Golden Rule transition events. As such, they satisfy a set of "intuitive" relations expressed by Eqs. \ref{eq:textbook}, \ref{eq:primitive}, \ref{eq:boltzmann}.

Our analysis of fourth-order corrections reveals interesting physics: the "intuitive" relations appear to be wrong, and corrections exhibit divergences in the limit of low temperatures manifesting an intriguing non-analytical dependence of the flows on coupling strength in the limit of weak couplings and vanishing temperatures.

\acknowledgments

The author is indebted to C.W.J. Beenakker, L.S. Levitov, K. Le Hur, I. Klich and F.W.J Hekking for numerous important discussions.


\begin{thebibliography}{99}
\bibitem{Definition} This differs from the popular definition ${\cal S}_M = \ln[{\rm Tr} \left\lbrace \hat{R}^M\right\rbrace]/(M-1)$ that would be extremely inconvenient to use in this Article.
\end{thebibliography}
\end{document}